\documentclass[11pt,preprint]{aastex}
\slugcomment{Submitted to the {\it Astrophysical Journal Letters}: Revised}
\def\fsecs{\hbox{$.\mkern-4mu^s$}}
\def\la{\mathrel{\hbox{\rlap{\hbox{\lower4pt\hbox{$\sim$}}}\hbox{$<$}}}}
\def\ga{\mathrel{\hbox{\rlap{\hbox{\lower4pt\hbox{$\sim$}}}\hbox{$>$}}}}

\shortauthors{Park}
\shorttitle{CXOU J160103.1}

\begin{document}

\title{Discovery of a Candidate Central Compact Object in the Galactic Nonthermal SNR G330.2+1.0}

\author{Sangwook Park\altaffilmark{1}, Koji Mori\altaffilmark{2}, 
Oleg Kargaltsev\altaffilmark{1}, Patrick O. Slane\altaffilmark{3}, 
John P. Hughes\altaffilmark{4}, David N. Burrows\altaffilmark{1}, 
Gordon P. Garmire\altaffilmark{1}, and George G. Pavlov\altaffilmark{1}} 

\altaffiltext{1}{Department of Astronomy and Astrophysics, Pennsylvania State
University, 525 Davey Laboratory, University Park, PA. 16802; park@astro.psu.edu}
\altaffiltext{2}{Department of Applied Physics, University of Miyazaki, 1-1 Gakuen 
Kibana-dai Nishi, Miyazaki, 889-2192, Japan} 
\altaffiltext{3}{Harvard-Smithsonian Center for Astrophysics, 60 Garden Street,
Cambridge, MA. 02138}
\altaffiltext{4}{Department of Physics and Astronomy, Rutgers University,
136 Frelinghuysen Road, Piscataway, NJ. 08854-8019}

\begin{abstract}

We report on the discovery of a pointlike source (CXOU J160103.1$-$513353)
at the center of a Galactic supernova remnant (SNR) G330.2+1.0 with {\it 
Chandra X-Ray Observatory}. The X-ray spectrum fits a black-body (BB) 
model with $kT$ $\sim$ 0.49 keV, implying a small emission region 
of $R$ $\sim$ 0.4 km at the distance of 5 kpc. The estimated X-ray luminosity 
is $L_X$ $\sim$ 1 $\times$ 10$^{33}$ ergs s$^{-1}$ in the 1 $-$ 10 keV
band. A power law model may also fit the observed spectrum, but the fit results
in a very large photon index, $\Gamma$ $\sim$ 5. We find no counterparts at other 
wavelengths. The X-ray emission was steady over the $\sim$13 hr observation
period, showing no variability. While we find marginal evidence for 
X-ray pulsations ($P$ $\approx$ 7.5 s), the presence of a pulsar at the 
position of this object is not conclusive with the current data, requiring
an independent confirmation. These results are generally consistent with an 
interpretation of this object as a Central Compact Object associated with SNR 
G330.2+1.0.  

\end{abstract}


\section {\label {sec:intro} INTRODUCTION}

Since the discovery of nonthermal X-ray emission from the shell of 
supernova (SN) 1006 \citep{koyama95}, the nonthermal X-ray emitting blast 
wave shock fronts of supernova remnants (SNRs) have been considered
as prime candidate sites for the cosmic-ray acceleration. Thanks to the 
high-resolution, high-sensitivity detectors on board {\it Chandra} and 
{\it XMM-Newton} observatories, several SNRs now show evidence for particle 
acceleration sites near the shock fronts. While the nonthermal X-ray 
emission is typically from localized regions near the shock front in young 
historical SNRs (e.g., Cas A, Tycho, SN 1006), nonthermal X-rays in some 
relatively ``older'' ($\tau$ $\ga$ a few 10$^3$ yr) SNRs dominate the entire 
face of the remnants \citep[G347.3$-$0.5 and G266.2$-$1.2,][]{koyama97,slane01}. 
Each of the latter also contains a Central Compact Object (CCO) 
\citep{slane99,pav01}, a peculiar manifestation of isolated neutron stars (NSs)
found in near the center of only a handful of SNRs \citep[e.g.,][]{pav04}.

Recently, using the archival {\it ASCA} data, Torii et al. (2006, T06 
hereafter) discovered that the X-ray emission from a Galactic shell-type 
radio SNR G330.2+1.0 \citep{caswell83} is dominated by nonthermal continuum
over the entire SNR with enhancements toward the boundary shell. 
The X-ray continuum was described by a power law (PL) model (photon index 
$\Gamma$ $\sim$ 2.8), indicating the X-ray synchrotron emission from the 
shock-accelerated electrons. Based on the distance of $\sim$5 $-$ 10 kpc 
\citep{mg01}, a Sedov age of $\tau$ $\ga$ 3000 yr was estimated, assuming 
the typical explosion energy ($E_0$ $\sim$ 10$^{51}$ ergs) and ambient density 
($n_0$ $\sim$ 1 cm$^{-3}$) (T06). G330.2+1.0 appears to resemble G347.3$-$0.5 
and G266.2$-$1.2, but, unlike those two SNRs, no central point source was 
found for G330.2+1.0 in the {\it ASCA} data (T06). 

We observed G330.2+1.0 with {\it Chandra} in order to perform a detailed
study of the SNR such as the spatially-resolved spectroscopy of the 
nonthermal X-ray emitting shell, which could not be performed with 
the low-resolution detectors of {\it ASCA}. The other major goal 
was the search for a NS at the center of the SNR, motivated by 
the fact that the other two nonthermal X-ray-dominated SNRs, G347.3$-$0.5 
and G266.2$-$1.2, contain a CCO. We report here the discovery of a pointlike 
source at the center of G330.2+1.0, which shows characteristics similar to the 
known CCOs. In this {\it Letter}, we present the results from the analysis 
of X-ray emission from this new candidate CCO. The results from the analysis 
of the SNR are presented elsewhere (Park et al., in preparation).

\section{\label{sec:obs} OBSERVATIONS \& DATA REDUCTION}

SNR G330.2+1.0 was observed with the Advanced CCD Imaging Spectrometer 
(ACIS) on board {\it Chandra X-Ray Observatory} in May 22, 2006 as part 
of the Guaranteed Time Observations program. G330.0+1.0 has an 
$\sim$10$^{\prime}$ angular size \citep[T06]{caswell83}. It is heavily 
absorbed and dominated by nonthermal continuum in the hard X-ray band 
($E$ = 1 $-$ 10 keV). We thus used the ACIS-I array to utilize its large 
field of view (FOV) of $\sim$17$^{\prime}$ $\times$ 17$^{\prime}$ and 
the relatively large collecting area in the hard band. The pointing was 
roughly toward the geometrical center of the SNR.  No strong variability 
was found in the background light curve. We corrected for the spatial and 
spectral degradation of the ACIS data caused by radiation damage, known 
as the charge transfer inefficiency (CTI; Townsley et al. 2000), with 
the methods developed by Townsley et al. (2002a), before further standard 
data screening by status, grade, and energy selections. ``Flaring'' pixels 
were removed, and the standard grades (02346) were selected. After the data 
reduction, the effective exposure was 49.5 ks. The overall SNR spectrum 
is hard, and there are few source photons below $E$ $\sim$ 1 keV. 
The central pointlike source, as discussed below, also shows a 
spectrum with nearly all source photons detected at $E$ $>$ 1 keV.
We thus extracted photons in the 1 $-$ 7 keV band for the data analysis.
(The upper bound of $E$ = 7 keV was chosen to avoid instrumental 
lines and the higher background rates at $E$ $>$ 7 keV.) 

\section{\label{sec:data} Data Analysis}

\begin{figure}[]
\figurenum{1}
\centerline{\includegraphics[angle=0,width=0.70\textwidth]{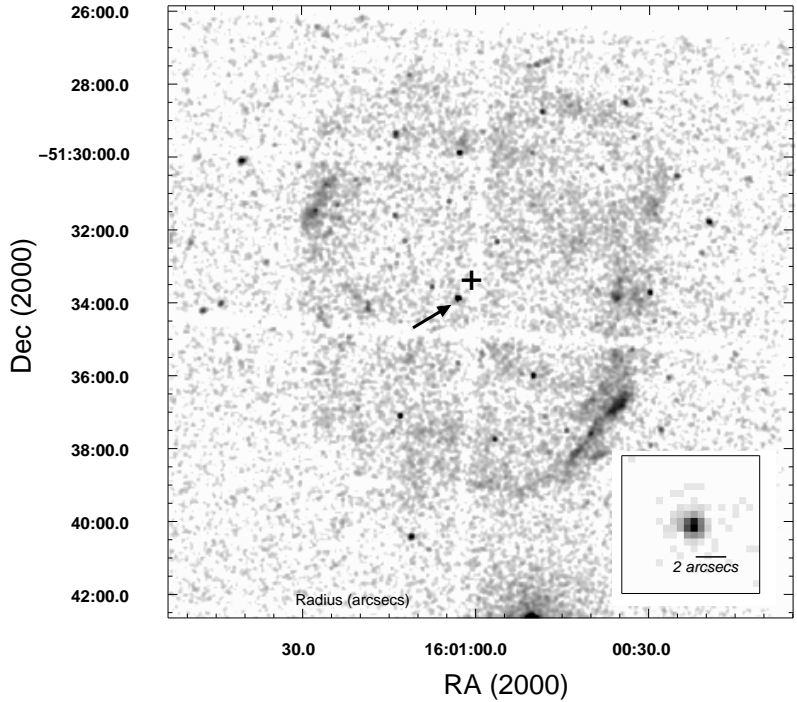}}
\figcaption[]{Broadband (1 $-$ 7 keV) ACIS image of G330.2+1.0.
Darker gray-scales correspond to higher intensities. For the purposes of display, 
the image has been binned with 4 pixels ($\sim$2$^{\prime\prime}$) and then 
smoothed by a Gaussian with $\sigma$ = 5 pixels. The NS candidate 
(CXOU J160103.1$-$513353) is marked with an arrow. The lower-right inset is 
the unbinned (0$\farcs$5 pixel) image of the NS candidate. The radio SNR center
\citep{caswell83} is marked with ``+''.
\label{fig:fig1}}
\end{figure}

The broadband image of SNR G330.2+1.0 is presented in Fig.~\ref{fig:fig1}. 
We detect some $\sim$20 pointlike sources within the SNR boundary 
($\sim$5$'$ radius). Among these sources, the one detected at the
``center'' of the SNR is remarkable. This source is the brightest 
($\sim$0.012 counts s$^{-1}$) among the sources detected within the SNR 
shell (all other sources are fainter by $\ga$ an order of magnitude.) 
The source position is nearly at the geometrical center of the SNR: 
e.g., only $\sim$33$^{\prime\prime}$ southeast of the radio SNR center
(Fig.~\ref{fig:fig1}) \citep{caswell83}. 
As we discuss below, this source appears to be a candidate NS associated 
with G330.2+1.0. Thus, we hereafter refer this object as 
the ``NS candidate''. We note that the NS candidate uniquely stands out in the 
central region of the SNR: e.g., within $\sim$1$'$ radius of the SNR center, 
we find only one other source at $\sim$45$^{\prime\prime}$ northeast of the 
NS candidate, which shows a soft spectrum and an optical counterpart and 
is thus most likely a foreground star. Comparisons of the NS candidate radial 
profile with the ACIS PSF at the source position show no evidence for an 
extended nebulosity, confirming the pointlike nature. The detected sky position 
of the source is RA(2000) = 16$^h$ 01$^m$ 3$\fsecs$14, Dec(2000) = 
$-$51$^{\circ}$ 33$^{\prime}$ 53$\farcs$6. It is detected within $\sim$1$'$ 
off-axis on the ACIS-I2, and thus the positional uncertainties are 
$\sim$0$\farcs$3 ($\sim$90\% C. L.) \citep{alex03}. We designate this source 
as CXOU J160103.1$-$513353. 

\begin{figure}[]
\figurenum{2}
\centerline{\includegraphics[angle=0,width=0.50\textwidth]{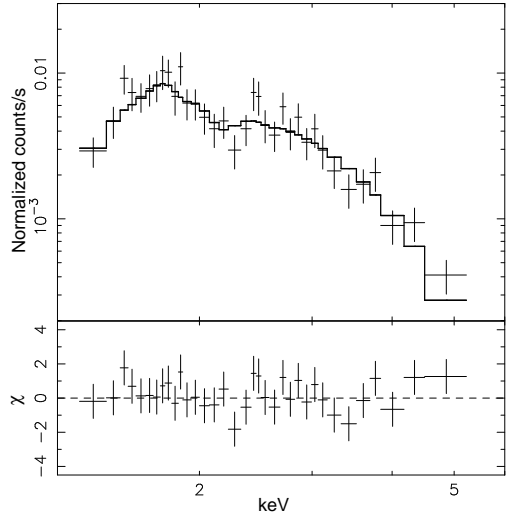}}
\figcaption[]{X-ray spectrum of the NS candidate as observed with the ACIS.
The best-fit BB model is overlaid. The lower panel is the residuals from 
the best-fit BB model.
\label{fig:fig2}}
\end{figure}

We extracted the source spectrum ($\sim$600 counts) from a circular 
region with a radius of 2$^{\prime\prime}$. The background spectrum 
was taken from a surrounding annular region (the inner radius = 
4$^{\prime\prime}$ and the outer radius = 15$^{\prime\prime}$). 
The source spectrum was binned to contain a minimum of 15 counts per 
energy bin. For the spectral analysis of our CTI-corrected data, we 
utilized the response matrices appropriate for the spectral 
redistribution of the CCD, as generated by Townsley et al. (2002b). 
The X-ray spectrum is continuum-dominated with no clear line 
features (Fig.~\ref{fig:fig2}). 
A single black-body (BB) model with $kT$ $\sim$ 0.49 keV can adequately 
describe the observed spectrum (Table~\ref{tbl:tab1}). The high 
foreground column ($N_H$ $\sim$ 2.5 $\times$ 10$^{22}$ cm$^{-2}$) is 
consistent with that for the SNR (T06, Park et al. in preparation). 
Alternatively, a PL model can fit the data, implying an extremely steep spectrum 
($\Gamma$ $\sim$ 5) and a higher $N_H$ $\sim$ 4.7 $\times$ 10$^{22}$ cm$^{-2}$. 
For completeness, we note that optically thin thermal plasma models 
($\tt mekal$) may also describe the spectrum with $kT$ $\sim$ 0.95 keV. 
The fit, however, requires an unrealistically low metal abundance with a 
90\% upper limit of $\sim$0.3 solar, which supports a continuum-dominated 
nature of the spectrum. 

\begin{figure}[]
\figurenum{3}
\centerline{\includegraphics[angle=0,width=0.50\textwidth]{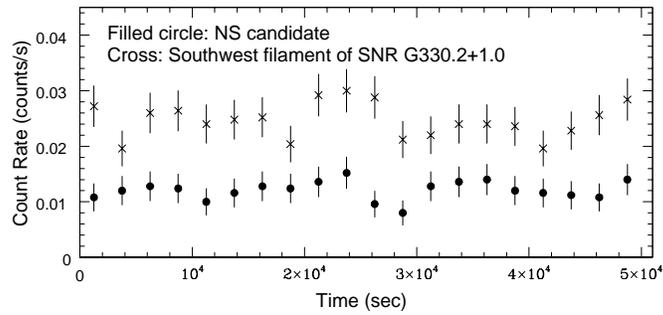}}
\figcaption[]{X-ray light curve of the NS candidate (filled circles).
For comparison, the light curve from the southwestern filament of SNR
G330.2+1.0 is presented with crosses. Each light curve has been binned
with 2500 s. 
\label{fig:fig3}}
\end{figure}

\begin{figure}[]
\figurenum{4}
\centerline{\includegraphics[angle=90,width=0.50\textwidth]{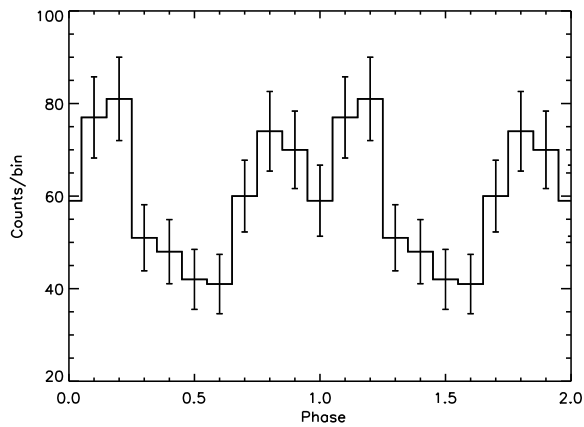}}
\figcaption[]{Pulse profile of the NS candidate folded with the frequency 
of 0.133663 Hz ($P$ = 7.48 s).
\label{fig:fig4}}
\end{figure}

The NS candidate shows no significant evidence of variability on time 
scales of $\sim$1 $-$ 7 hr (Fig.~\ref{fig:fig3}). Although our {\it Chandra}
observation was perfomed with the standard 3.24 s frame-time, which
precludes the search for an ordinary pulsar-like periodicity ($P$ $\la$ 1 s), 
we can attempt to search for a long periodicity of $P$ $\sim$ several seconds, 
typical for Anomalous X-ray Pulsars (AXPs) \citep{mere02}.
The ACIS time resolution of 3.24 s and the total exposure of $\sim$50 ks 
allow us to search for pulsations in the $4\times 10^{-5}$ $-$ 0.15 Hz range. 
We use the arrival times of the 603 photons (of which $\sim$$99.8$\% are 
expected to come from the source), recalculated to the solar system barycenter 
using the {\tt axBary} tool. We used the $Z_{m}^{2}$ test \citep{buc83} to 
search for periodic pulsations. We calculated $Z_{m}^{2}$ for $m=1$ $-$ 4 (where
$m$ is the number of harmonics included) at $1.5\times10^{5}$ equally spaced 
frequencies $f$ in the $4\times 10^{-5}$ $-$ 0.15 Hz range. This corresponds to 
oversampling by a factor of about 20, compared to the expected width of 
$T_{\rm span}^{-1}\approx 20$ $\mu$Hz of the $Z_{m}^{2}(f)$ peaks, and guarantees 
that we miss no peaks. The most significant peak was found in $Z_{2}^{2}$ (= 29.5,
corresponding to the $\approx$95\% significance level) at $f=0.133663\, {\rm Hz}\pm 5\, 
\mu{\rm Hz}$ ($P\approx 7.48$ s) (Fig.~\ref{fig:fig4}). 
If this corresponds to the actual period of the source, the pulsed fraction 
is $\approx$$30\%$. However, we note that the statistical significance of the period is 
only marginal. If this period is simply due to the statistical fluctuation,
we place a 3$\sigma$ upper limit of $\approx$52\% on the pulsed fraction, 
assuming sinusoidal pulse shape per period. Follow-up 
observations with higher time resolution and better photon statistics will 
be essential to verify the suggested periodicity or to search for a
different period. 

\section{\label{sec:disc} Discussion}

The sky position at the center of SNR G330.2+1.0 and the high foreground 
column, which is consistent with that of the SNR, make the NS candidate 
association with the SNR highly plausible. There is no evidence for 
hour-scale variability that might be caused by orbital and/or stellar 
activities. We find no counterparts at other wavelengths. There is no radio 
(843 and 1415 MHz) enhancement at the SNR center \cite{caswell83}, and 
the nearest cataloged radio pulsar (PSR B1558$-$50) is $\sim$36$'$ away.
The nearest optical source (USNO B1.0 \#0384$-$0529508, B = 16.92, R = 15.08)
is $\sim$3$\farcs$7 off the NS candidate. Considering the accuracy
of both the X-ray and the optical positions ($\la$0$\farcs$3), this source 
is unlikely to be the optical counterpart of the NS candidate. The optical 
upper limit (V $\approx$ 21) then implies a large X-ray-to-optical flux ratio
of ${f_{X(1-10~{\rm keV})}}/{f_V}$ $>$ 9, which supports the NS interpretation
rather than a normal stellar or extragalactic object \citep{stocke91,agu06}.
The high absorbing column toward G330.2+1.0 supports previously estimated
distances of $d$ $\sim$ 5 $-$ 10 kpc. 
Assuming that the NS candidate is associated with G330.2+1.0, we scale the
distance to the NS candidate in units of 5 kpc.

The best-fit BB temperature of the NS candidate is higher than those
expected from the standard cooling of the young NS's surface \citep{yak04},
but it is consistent with the temperature observed in CCOs and/or AXPs 
\citep[2004]{mere02,pav02}. The small size of the implied BB radiating region, 
$R_{BB}$ $\sim$ 0.4 $d_5$ km (where $d_5$ is the distance scaled to 5 kpc), 
is similar to that of other CCOs \citep[Pavlov et al. 2000,2002;][]{kar02}, 
probably indicating the X-ray emission from a hot polar cap of the NS. 
The X-ray luminosity $L_{X(1-10~{\rm keV})}$ $\sim$ 1 $\times$ 10$^{33}$ 
$d^2_5$ ergs s$^{-1}$ is also consistent with the object being a CCO (Pavlov 
et al. 2004 and references therein). The PL model implies a higher $N_H$ than 
in the BB case and a very steep slope of the spectrum. These parameters
resemble those of the CCOs in the other two nonthermal X-ray SNRs G347.3$-$0.5 
and G266.2$-$1.2 \citep{kar02,laz03}. The higher column associated with the PL 
fit could suggest a background AGN origin of the NS candidate, as perhaps suggested 
by the relatively high chance probability ($\sim$0.2) of such a detection within 
the SNR boundary ($\sim$5$'$ radius) based on the LogN-LogS relation in the 
Galactic plane \citep{ebi01}. However, the combination of the extremely steep 
X-ray spectrum ($\Gamma$ $\sim$ 5), the high ${f_X}/{f_O}$ ratio, the absence 
of a radio counterpart, and the suggested possible pulsations 
strongly favor the case of a NS, making an AGN interpretation highly unlikely.

The NS candidate was not detected with {\it ASCA} (T06). 
The non-detection might be attributed to the low sensitivity 
(e.g., the GIS count rate is $\sim$30\% of the ACIS-I at $E$ $>$ 1 keV) 
and the poor angular resolution ($\sim$3$^{\prime}$ FWHM) of {\it ASCA}. 
In fact, the $\sim$20 ks exposure of the archival {\it ASCA} 
data implied only $\sim$70 photons per GIS over the $\ga$10 arcmin$^2$ 
source area. 
On the other hand, we serendipitously detected the NS candidate in 
archival {\it XMM-Newton} data (ObsID 0201500101). Assuming a BB model with 
$kT$ = 0.5 keV and $N_H$ = 2.5 $\times$ 10$^{22}$ cm$^{-2}$, we 
estimate $L_X$ $\sim$ 1.3 $\times$ 10$^{33}$ ergs s$^{-1}$ for the 
{\it XMM-Newton} source\footnote{The short exposure of $\sim$8 ks provided
only limited photon statistics ($\sim$50 counts per MOS) and did not 
allow further analysis. It was off the pn FOV due to the use of the 
small window mode.}, which is consistent with our {\it Chandra} results. 

Based on the overall spectral/temporal properties and the position at the 
center of SNR G330.2+1.0, we propose that CXOU J160103.1$-$513353 is a candidate
CCO associated with G330.2+1.0. The CCOs are believed to be a peculiar 
manifestation of NSs, whose nature is not fully understood. Considering only 
a handful of the known CCOs, the discovery of this new candidate CCO is 
important as it adds a new member to this class of NSs. The association with 
a newly-discovered nonthermal SNR G330.2+1.0 makes this new CCO candidate 
particularly interesting because each of the other two nonthermal 
X-ray-dominating SNRs also contains a CCO. Furthermore, the possible 
pulsations at $P$ $\sim$ 7.5 s (if confirmed), which is typical for AXPs, 
makes this object potentially even more intriguing, because only two of the 
known CCOs show pulsations, both with shorter periods \citep{zav00,got05}. 
If the pulsations with the long period are confirmed with follow-up observations, 
the NS candidate may considered to represent a ``bridging'' class between 
CCOs and AXPs.  

\acknowledgments

This work was supported in part by SAO under {\it Chandra} grant SV4-74018.
OK and GGP were supported by NASA grant NAG5-10865. KM was partially
supported by the Grant in-Aid for Young Scientists (B) of the MEXT (No.
18740108).

\clearpage

\begin{deluxetable}{lccccccc}
\tabletypesize{\footnotesize}
\tablecaption{Best-Fit Spectral Parameters of the NS Candidate.
\label{tbl:tab1}}
\tablewidth{0pt}
\tablehead{\colhead{Model} & \colhead{$N_H$} & \colhead{$\Gamma$} & 
\colhead{$kT$} & \colhead{$f_{X(1-7~{\rm keV})}$\tablenotemark{a}} & 
\colhead{$L_{X(1-10~{\rm keV})}$} & \colhead{$R_{BB}$\tablenotemark{b}} & 
\colhead{$\chi^2$/$\nu$}  \\ 
& \colhead{10$^{22}$ cm$^{-2}$} & & \colhead{keV} & 
\colhead{10$^{-13}$ ergs cm$^{-2}$ s$^{-1}$} & \colhead{10$^{33}$ $d^2_5$ 
ergs s$^{-1}$} & \colhead{$d_5$ km} & }
\startdata
BB & 2.5$^{+0.5}_{-0.4}$ & - & 0.49$^{+0.04}_{-0.04}$ & 1.17$\pm$0.10 & 
1.0 & 0.40$^{+0.34}_{-0.17}$ & 28.4/33 \\
PL & 4.7$^{+0.7}_{-0.6}$ & 5.0$^{+0.6}_{-0.5}$ & - & 1.23$\pm$0.10 & 
6.3 & - & 25.0/33 \\
\enddata
\vspace{-5mm}
\tablecomments{Errors are with 90\% confidence.}
\tablenotetext{a}{Obeserved X-ray flux.}
\tablenotetext{b}{Black-body radius.}
\end{deluxetable}


\end{document}